\documentclass{aa}

\usepackage{natbib} 
\usepackage{psfig} 

\begin{document} 
 
\title{On the nature of bimodal initial 
velocity distribution  of neutron stars} 
 
\titlerunning{On the bimodal kick velocity distribution} 
 
\author{I. Bombaci 
         \inst{1,2} 
          \and 
S.B. Popov 
            \inst{3,4} 
         } 
 
   \offprints{S. Popov} 
 
\institute{INFN Sezione di Pisa, 56127 Pisa, Italy 
               \and 
Universit\`a di Pisa, Dipartimento di Fisica, 
via F. Buonarroti 2, 56127 Pisa, Italy\\ 
\email{bombaci@df.unipi.it} 
              \and 
Universit\`a di Padova, Dipartimento di Fisica,  
via Marzolo 8, 35131, Padova, Italy 
            \and 
   Sternberg Astronomical Institute, 
Universitetski pr. 13, 119992 Moscow, Russia\\ 
\email{polar@sai.msu.ru} 
}

   \date{} 
 
\abstract{ 
We propose that the bimodal nature of the kick velocity distribution   
of radio pulsars is connected with the  
dichotomy between  
hadronic stars ({\it i.e.} neutron stars with no quark matter content)  
and quark stars.     
Bimodality can appear due to different mechanisms of explosion   
which leads to the  formation of  two types of compact stars 
or due to two different sets of parameters mastering a particular 
kick mechanism.    
The low velocity maximum (at $\sim 100$~km~s$^{-1}$) is  
connected with hadronic  star formation, 
whereas the second peak corresponds to quark stars.  
In the model of delayed collapse of hadronic stars to quark stars 
(Berezhiani et al. 2003\nocite{bbd2003})  quark deconfinement  leads  
to a second energy release,  and  to a second kick, 
in addition to the kick imparted to the newly formed hadronic star 
during the supernova explosion.    
If the electromagnetic rocket mechanism can
give a significant contribution to pulsar kicks,
then the high velocity peak can be connected with the shorter initial 
spin periods  of quark stars with respect to hadronic stars.      
We discuss {\it pro et contra} of these scenarios.  
} 
 
\maketitle 

\pagebreak
 
\section{Introduction} 
 
 An important property of the compact stars usually called   
{\it neutron stars}  (NSs) is their spatial velocity distribution.    
 At present there are many measurements of radio pulsars (PSRs)  
proper motions  (Brisken et al. 2003 and references there in).   
These measurements, together with distance measurements,   
lead to the determination of pulsar spatial velocities.     
In the ATNF catalogue  (see for example Hobbs et al. 2003\nocite{hobbs2003})   
there is data on transverse velocities for 137 PSRs  
(including millisecond etc.).     
Typical  velocities of PSRs are about 300-400~km~s$^{-1}$ (Lyne \& Lorimer   
1994\nocite{ll1994}, Lorimer et al. 1997\nocite{lbh1997}).    
The largest inferred  velocities are higher than  1000~km~s$^{-1}$.   
   
Such high pulsar velocities are thought to be connected with supernova (SN)  
explosions.   
In fact, it is believed that a newly born neutron star receives a considerable 
``kick'' during, or shortly after, the SN explosion  
(see reviews in Lai  2003\nocite{lai2002},  Burrows et al. 2003\nocite{bom2003}).   
Understanding of the nature of pulsar velocity distribution can give 
important information about the  physics of SN.   
  
There have been various attempts to reconstruct the initial velocity  
distribution starting from  the observed properties of radio pulsars and  
other NSs (for example NSs in SN remnants -- SNR; NSs in close binaries  
etc.).    
At the present time the most widely accepted velocity distribution    
is the one  obtained by \cite{acc2002}.  It is a bimodal distribution with  
two Maxwellian components (see Fig.~1). 
The velocity dispersion, $\sigma$,  of the first peak is  90~km~s$^{-1}$.   
The fraction of  compact objects in the first peak is  about $0.4\pm 0.2$.  
The second peak corresponds to $\sigma=500$~km~s$^{-1}$.   
Recent results by \cite{bfg2003} also support a two-component 
velocity distribution (however, in their model the fraction of the   
low-velocity component is just 0.2).   
 
In principle the initial velocity distribution can be multi-component as  
far as there are many potential sources for different velocity maxima   
(binaries of different types, different types of SNae etc.),  but the relative  
contribution of many of them should be small (for example,   
the contribution of accretion induced collapse (AIC) or possible  
compact remnants of faint SNIa etc.).  
In addition compact stars,  which never appear as PSRs, can have  
a velocity distribution different from the PSRs' one.     
In the present paper,  we will concentrate on the spatial velocity   
distribution for radio pulsars deduced from observational data,   
and we will try to give a physical interpretation of the origin   
of its  bimodal shape.     
 
There have been several discussions connected with  this  bimodality
or with different possible components of the kick velocity distriburion.  
For example \cite{cw2002} suggested a modification of the \cite{i1992}  
SN mechanism (see also Imshennik \& Nadezhin 1992\nocite{in1992}) 
to explain high-velocity part of the distribution.   
This mechanism (as noted by the authors) is not valid for low-velocity NSs  
as Crab and Vela, for which the spin is nearly aligned with the spatial  
velocity (see recent data and discussion in Romani 2004\nocite{r2004}).  
Actually, such a geometrical configuration is not possible 
in the mechanism discussed by Colpi and  Wasserman.        
\cite{plp2003} proposed the possibility of  low kick velocities  
for a particular type of high-mass X-ray binaries 
(HMXBs) as a result of mass exchange influence  
on the stellar evolution.   
   
In the present paper we propose a different scenario connected  
with the occurrence of quark deconfinement in the cores of neutron stars,    
and with the corresponding energy release.   
   
\begin{figure}
\vbox{\psfig{figure=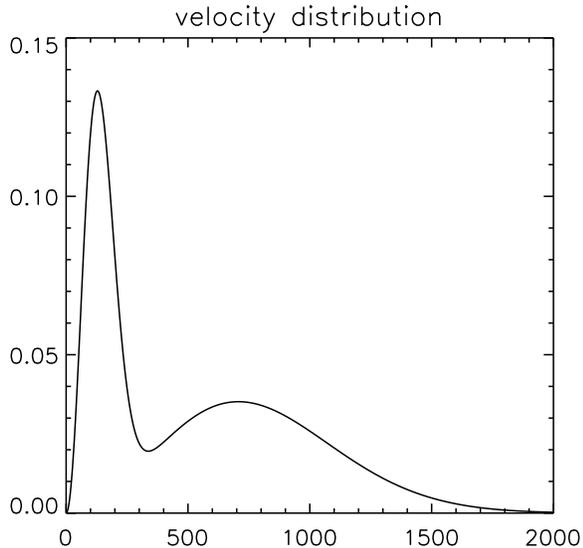,width=\hsize}}  
\caption[]{ The initial velocity distribution 
with two maxwellians (Arzoumanian et al. 2002\nocite{acc2002}).   
The low velocity part includes 40\% of all objects, the maximum corresponds to 
127~km~s$^{-1}$. The high velocity part has maximum at 707~km~s$^{-1}$. 
} 
\label{fig:velo} 
\end{figure} 


\section{The Model}  
\label{model}  
 
In the following we assume that the initial velocity distribution  
of radio pulsars is at least bimodal,  and fractions of stars in the low and  
high velocity parts of the distribution are more or less equal.   
It means that among normal radio pulsars there should be at least   
two different types of objects.  
 
The fact that we are speaking about radio pulsars tells us that to produce  
different kicks we cannot use ideas which involve physical conditions  
for which radio pulsars do not exist (for example, too low magnetic fields,  
or too long initial rotational periods).   
Binaries also cannot explain this bimodality (they do not produce enough  
high-velocity NSs if isolated and binary progenitor stars produce PSRs with
comparable probabilities).   
So, one has to think about some significant dichotomy among normal  
single (isolated) radio pulsars.    
 
The strongest theoretically known dichotomy among compact stars  
is the dichotomy between pure {\it hadronic stars} (HSs)  and  
{\it quark stars} (QSs) 
(see {\it e.g.}  Glendenning 1996\nocite{gle96}, Bombaci et al. 2004\nocite{bpv2004}).    
The physical feature which distinguishes the two families of compact stars  
is the absence or the presence of  deconfined quark matter in the  
stellar interior.  In hadronic stars no fraction of quark matter is  
present. These stars below the usual stellar crust 
(Pethick \&  Ravenhall 1995\nocite{pr95})     
have a layer  of neutron-rich nuclear matter in beta-equilibrium with   
electrons and muons,  
and possibly an inner core  containing hyperons ({\it hyperon stars})  
or a condensate of negative kaons in addition to the particles mentioned above.     
Compact stars which possess  a quark matter core,  either as a mixed phase  
of deconfined quarks and hadrons or as a pure quark matter phase, are called  
{\it hybrid stars} (HySs).     
A more exotic alternative to the existence of hybrid stars is the possible  
existence of self-bound compact stars consisting completely of a  
deconfined mixture of {\it up} ({\it u}),  {\it down} ({\it d})  and  {\it strange } ({\it s})  
quarks  (together with an appropriate number of electrons to guarantee electrical  
neutrality)  satisfying the Bodmer--Witten hypothesis   
(Bodmer 1971\nocite{bod71}; Witten 1984\nocite{witt84}; 
see also Terazawa 1979\nocite{ter79}).   
These  compact stars are called  {\it strange stars} (SSs)  
(Alcock et al. 1986; Haensel et al. 1986\nocite{hae86}).       
In the following, we will refer  to hybrid stars  and  strange stars  
collectively as QSs.    
 
Compact stars of each of the two classes ({\it i.e.} hadronic stars and  
quark stars) could be endowed with strong magnetic fields and could  
manifest their presence in the Universe as pulsars or as compact X-ray  
sources in binary systems.  

Among many differences between HSs and QSs, 
typical kick velocities for the members of these two families 
of compact stars can be different too.     
It is a surprising fact that nobody (in our  knowledge) discussed 
in details the possibility  
of having different kick velocities for these two types of compact objects  
except a short note by \cite{zxq2000}.   
These authors briefly mentioned, in connection with larger displacement  
of soft gamma-repeators (SGRs) in comparison with anomalous X-ray pulsars
(AXPs) from the centers of their SNRs,  that strange  stars can  
have higher kick velocities due to a two-step kick mechanism.   

\begin{figure}
\vbox{\psfig{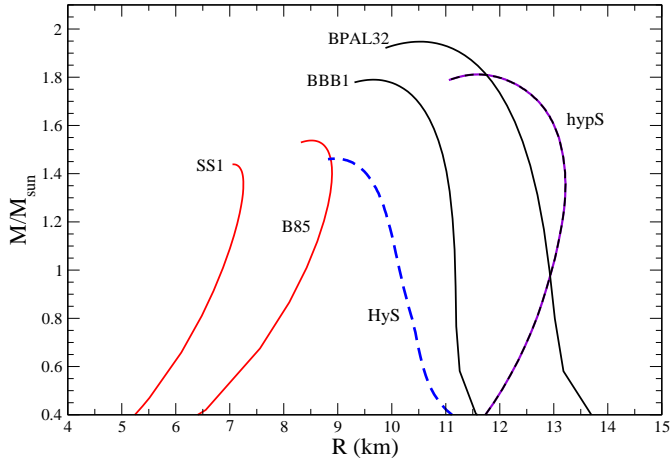}}  
\caption[]{ The mass-radius relation for different types of compact stars.  
The curves labeled BBB1 (Baldo et al. 1997\nocite{bbb97}) and  
BPAL32 (Prakash et al. 1997\nocite{pra97}) 
are related to pure nucleonic compact stars ({\it i.e.} hadronic stars 
with the whole core made of $\beta$-stable nuclear matter), 
the curve labeled with hypS --- to hyperonic stars (Glendenning 1992\nocite{gle92}).  
The curve labeled with HyS  is relative to a hybrid star 
(Bombaci et al. 2004\nocite{bpv2004}).  
Finally, the two curves  SS1 (Dey et al 1998\nocite{dey98})  and B85 are related 
to strange stars.   
Stellar masses are plotted in unit of the solar mass.
} 
\label{fig:MR} 
\end{figure} 

A quark star  can be formed in  a delayed collapse   
(see Berezhiani et al. 2003\nocite{bbd2003})   
or in a direct collapse with additional energy release  
(see for example Benvenuto \& Horvath 1989,  
Hong et al. 2001\nocite{hhs2001}).   
Let us first discuss the delayed collapse scenario.   
 
It has been recently shown (Berezhiani et al. 2003\nocite{bbd2003}, 
Bombaci et al. 2004\nocite{bpv2004})  
that above a threshold value of the gravitational mass  pure hadronic stars  
are metastable to the conversion into quark stars  
(hereafter the HS$\rightarrow$QS conversion).       
The  {\it mean-life time}  of a metastable hadronic star is related to the  
quantum nucleation time to form a drop of quark matter in the stellar center,   
and  dramatically depends on the value of the stellar central pressure.   
Thus, a metastable  hadronic star can have a mean-life time which can  
span many orders of magnitude (Berezhiani et al.  2003\nocite{bbd2003}, 
Bombaci et al. 2004\nocite{bpv2004}).  
In other words, the HS mean-life time could be many orders of magnitude  
larger than the age of the Universe,  or on the opposite side  
it can be relatively short: from a few hours up to a few years.   
 
 
The delayed collapse of a  metastable hadronic star to a quark star  
may be triggered by a  fall-back ({\it i.e.} relatively 
soon after the SN explosion forming the HS)  
or by accretion of matter in a binary system.        
Both of these processes  increase the central stellar density and,   
consequently,  strongly decrease the nucleation time to form  
the  first  critical-size drop of quark matter which starts the stellar  
conversion process.    
As  we suggest that about one half of all PSRs are QSs,  
then {\it most} of QSs should be formed due to fall-back or directly  
in the SN explosion (so that the delay time is significantly shortened),   
since the outcome  of NSs from binary systems  
is rather small to constitute $\sim$1/2 of the total PSR population.   
Anyway,  an additional energy release can produce  a bimodal  
distribution of many physical properties  of the compact stars  
which we usually call just NSs.

We suggest that the delayed stellar conversion process of a pure HS to a QS,    
triggered by quark deconfinement,   can lead to a  second kick 
to the  compact star.   
Thus, the low velocity component of the pulsar velocity distribution 
receives contribution mainly from ``normal'' NSs  (hadronic stars) 
which passed through a single  explosion (the SN explosion).  
The high velocity part is mostly composed of QSs which received 
a second kick  due to the energy release associated with the  
stellar conversion process.   
 
Of course, as far as both explosions produce {\it distributions}  
of velocities, it is possible to find low velocity QSs which passed through   
two kicks, and high velocity HSs which received  just a single natal kick.     
Also,   among mid- and high-velocity objects  there can be a contribution  
of HSs from massive binaries  
(Iben \& Tutukov 1996\nocite{it1996}). 
 
\begin{table*}  
\caption{Kick mechanisms}  
\begin{tabular}{|lcccc|}  
\hline 
   & & & & \\ 
Mechanism & Time scale & $V_\mathrm{max}$, & Alignment & Main recent refs.  \\ 
   & & km~s$^{-1}$ & (spin and $V$) & \\                                       
\hline                                                   		  
   & & & &\\ 
Hydrodynamical & 0.1 s & $\sim(100-200)$& random & \cite{lcc2001}  \\ 
   & & & & \\ 
$\nu$-driven & $\sim$ few s& $\sim$ 50~$B_{15}$ & parallel & \cite{lcc2001}  \\ 
   & & & & \\ 
Electromagnetic rocket & long & 1400~$R_{10}^2\, P_\mathrm{ms}^{-2}$ &  
parallel & 
\cite{lcc2001},  \\ 
   & & & & \cite{hdl2003} \\ 
Binary disruption & $<<P_\mathrm{orb}$& $\sim$ 1000 & perpendicular & \cite{it1996}  \\ 
(without add. kick) & & & & \\ 
NS instability & few ms &  $\sim$ 1000 & perpendicular & \cite{cw2002},  \\                                       
   & & & & \cite{ir2004}  \\           
Magnetorotational & 0.2 s --  & $\sim 300$ & quasirandom  & \cite{mba2003}, \\ 
                  &  minutes  & (up to 1000) &   & \cite{abm2003}\\ 
\hline  
\end{tabular} 
\end{table*} 

Kicks can be imparted to compact stars due to several different  
mechanisms (see table 1).    
Let us estimate the additional spatial velocity imparted to a QS formed via  
the delayed HS$\rightarrow$QS conversion process 
in the case the kick is the result  
of  two asymmetric  neutrino jets.   
The total energy, $E_{\mathrm{Total}}$,  liberated in the stellar conversion  
is about a few $10^{53}$ ergs (Bombaci \& Datta 2000, Berezhiani et al.
2003\nocite{bbd2003}).   
A fraction $\eta$ of this energy  goes to a ultra-relativistic beamed   
$e^+e^-$ plasma.  The {\it efficiency factor} $\eta$  is related to the cross section  
for the process of neutrino-antineutrino annihilation into $e^+e^-$ pairs.   
Near the surface of the nascent QS due to general relativity effects  
the efficiency of the $\nu{\bar\nu}\rightarrow e^+e^-$  process is strongly  
enhanced with respect to the   Newtonian case (Salmonson \& Wilson 1999),   
thus $\eta$ can be as high as 0.1.   
To be more conservative let us assume $\eta = 0.01$-$0.1$.  
Next we assume that  the asymmetry between the two jets, $\kappa$,   
can vary from a few percent up to ten(s) percent.    
So the energy unbalance between the two jets is:   
 
$$   
\Delta E=\eta \kappa E_\mathrm{Total} \sim {\rm  few} \,
10^{49} - 10^{51} \mathrm{erg}   
$$   
for $\eta$ and $\kappa$ varying between 0.01 and 0.1.  
The associated momentum difference is:  
  
$$  
       \Delta P = \Delta E/c.  
$$  
  
Total moment conservation then requires for the kick velocity:  
$$ 
V_\mathrm{NS}/c = \Delta E/(M_\mathrm{NS}c^2)  
$$ 
So finally we  obtain $V_\mathrm{NS}\approx$ ~ few (1--100) km~s$^{-1}$.    
 
A similar hydrodynamical neutrino driven kick mechanism could give  
the first kick to the HS during the SN explosion (Goldreich et al. 2000).

The second possibility for a QS to obtain a larger kick velocity 
with respect to an HS is the following.  
If compact star kicks are produced by the electromagnetic rocket 
mechanism (Harrison \& Tademaru 1975\nocite{ht1975}),  
then the high velocity peak can be connected  with  
a shorter initial stellar rotational period of QSs with respect to that of HSs.   
Of course, this mechanism is valid both for QSs formed directly in 
the SN explosions  (Benvenuto \& Horvath 1989, Hong et al. 2001\nocite{hhs2001}),     
and for QSs formed with significant time delay (Berezhiani et al. 2003\nocite{bbd2003}). 

 Within the electromagnetic rocket mechanism  the maximum  
stellar velocity can be estimated as (Lai et al. 2001\nocite{lcc2001}):  
  
\begin{equation}  
V_\mathrm{max} \sim 1400 R_{10}^2 P_\mathrm{ms}^{-2}$~km~s$^{-1}  
\label{vmax}  
\end{equation}  
where $P_\mathrm{ms}$ is the stellar rotational period in milliseconds,  
and $R_{10}$ is the radius of the star in unit of 10 km.

Radii of QSs are typically 1.5--2  times lower than radii of  HSs.   
This is illustrated in Fig. 2, where we plot the mass-radius relation for  
different types of  compact stars (see figure caption for more details 
and references  to the stellar models).      
If  the initial rotational periods of QSs  are 4--5 times shorter 
than those for HSs,  then we can expect spatial velocities 5-6 times larger .     
Such a difference would be sufficient to explain the difference 
between the two peaks  in the initial velocity distribution.     

In the model of the HS$\rightarrow$QS conversion,  
it is difficult to increase the stellar spin  frequency by a factor of 4 in 
the process of  HS contraction into a QS,    
unless there is some additional spin-up due to the kick (see below).   
However, if a QS is formed during the SN explosion 
({\it i.e.} without time delay),   then the  difference in the initial  
periods between QSs and HSs could  be significant.  
In fact, different types of compact stars can suffer diverse instabilities 
which lead to  spin-down, depending on the stellar  composition.   
For example, strange stars  are not subject to the r-mode instability   
(Madsen 1998\nocite{m1998}).    
Typical initial periods of ``normal'' NSs (HSs) are about 20~ms or longer.   
With such 
values for the spin period, the electromagnetic rocket mechanism 
is not effective.     
If QSs have typical radii of about 7~km and if they are born with typical 
periods of about 1~ms,   then they can form a peak around 700~km~s$^{-1}$.   
\cite{whw2002} note that compact stars are expected to be born  
with periods $\sim 1$~ms, and only due to r-mode instability 
or similar  processes they are expected to slow down to rotational periods 
larger than about 20~ms.    

As it was discussed in \cite{sp1998} and in  \cite{pp1998} the kick 
itself can spin-up a newborn NS. So, in several kick-mechanisms 
there is a possibility of higher initial spin frequency for higher initial kicks. 
Such a situation is in favour of having higher velocities  
due to electromagnetic  
rocket mechanism if (before it starts to be active) the spin period had been
decreased by a kick. In that case even a moderate electromagnetic-rocket
kick can be important as far as it would be active for stars with already
significant kick. In addition non-central first kick can 
probably lead to dipole
displacement which causes more efficient acceleration dut to electromagnetic
rocket mechanism.
     
In principle, if an exact shape of the second velocity peak is known  
then for the electromagnetic rocket mechanism it is  
possible to reconstruct the form of QSs spin period distribution.  
But at the moment there is no data on the exact shape. Maxwellian   
distribution is just a reasonable  
hypothesis used by \cite{acc2002}.  Reconstruction of the initial  
velocity distribution (especially on the high end) basing on the radio   
pulsar proper motion data cannot distinguish between Maxwellian and similar  
shapes.


 
\section{Discussion} 
 
%

\subsection{Problems of the scenarios}

Here we discuss some possible problems of both scenarios   
(delayed HS$\rightarrow$QS conversion and electromagnetic rocket) 
for stellar kicks, paying more attention to the former one.  

One problem for a delayed HS$\rightarrow$QS conversion is connected 
with the surviving of  close binaries, where the compact star has  accreted 
a significant amount  of matter which could lead to quark deconfinement 
and subsequent  HS$\rightarrow$QS conversion.    
If most of the binaries do not survive the stellar conversion, 
due to the additional kick imparted to the newformed QSs,   
it will be  necessary to claim that all the observed  
accreting compact stars  in low-mass X-ray binaries (LMXBs) 
are HSs  
{\footnote{Compact objects in HMXBs with moderate kicks discussed by  
\cite{plp2003} also should be normal NSs according to our scenario.}},   
or it will be necessary to claim that in the case  
of binaries the HS$\rightarrow$QS conversion  does not lead to a large recoil.   
However, even with a strong kick there is a non-negligible probability  
of binary survival.  
Cir X-1  which has high spatial velocity 
(see discussion in Lai et al. 2001\nocite{lcc2001})   
can be an example. 
 Thus,  the compact star SAX~J1808.4-3658 (a member of a LMXB) 
could be a strange star as suggested by Li et al.  (1999\nocite{li99a}).  

We note that in the electromagnetic rocket model,  binaries can survive as 
far as the time of acceleration is long, unless the NS is a magnetar  
(Lai et al. 2001\nocite{lcc2001}, Huang et al. 2003\nocite{hdl2003}).   
This mechanism can lead to the formation of runaway binaries with QSs.  
 
Another problem comes from millisecond PSR (mPSR) velocity distribution. 
These objects are assumed to originate in LMXBs 
or in intermediate-mass X-ray binaries (IMXBs) where they accreted a 
significant amount of matter, so in the scenario of \cite{bbd2003}, 
one can expect that  most of mPSRs are QSs. 
As it was shown by \cite{lyne1998},  the  average velocity of mPSR 
is rather low: $\sim 130$~km~s$^{-1}$ 
(or even lower, see Nicastro et al. 2001\nocite{n2001}).  
This is consistent with investigations of binary evolution ({\it i.e.} no
additional kick is necessary). 
After the SN explosion,  most of the survived binaries correspond 
to the low velocity part of kick velocity distribution.    
Thus,  in our scenario, most of the compact stars in 
LMXBs are more likely HSs, not QSs, 
since the  mPSR population does not show any signs of enhanced velocities 
(there is a higher probability, that a QS can stay bound in a HMXB system).   
 
If the idea of the second kick due to the delayed HS$\rightarrow$QS 
conversion  is proved to be in contradiction with observations
({\it i.e.} no "imprints" of an additional kick on orbital parameters or/and
velocity distribution are found),  
then it is a serious problem for the whole scenario 
of delayed HS$\rightarrow$QS conversion  due to accretion in binaries 
as far as it is difficult to understand the  absolute  
symmetry of an energy release of $\sim 10^{53}$~erg~s$^{-1}$ from a compact 
object which is not a black hole.  
However, the physical conditions for the second collapse   
can be different in the case of accretion in a binary and in  
the case of a fall-back.  
In the first case a  
pre-collapsed object is a millisecond NS with low magnetic 
field ($\sim10^9$~G). In the second case there is a protoNS with longer spin 
period and higher magnetic field. This can lead to different asymmetries of 
explosions.  

The rocket scenario also has some problems.  
The main is that we need to assume significant displacement 
of the dipole relative to the spin axis. Otherwise the coefficient in  
eq.\ref{vmax} 
is much smaller and it is necessary to assume very short periods together 
with relatively large radii to explain PSRs with $V>1000$~km~s$^{-1}$. 


\subsection{Supernovae} 
 
At the moment nearly 3000 SNae have been  observed. 
Potentially different types of SNae can give different velocity distributions 
of the  resulting compact objects (if they are formed). 
So, as it was mentioned above, the initial velocity distribution should have 
many components. Different components can be associated with different SNae. 
It is very difficult to check observationally which SNae produce different 
compact objects. Most of the know SNae are extragalactic and recent  
-- for them there is good data about types (and even data on SN 
progenitors!, see Van Dyk et al. 2003\nocite{vlf2002}), but 
no  information about compact objects. For historical SNae (Green \& 
Stephenson 2003\nocite{gs2003})  
we often know a compact object, but it is very difficult to 
reconstruct the SN type. Then there are SNRs associated with compact objects 
(see Kaspi 2000\nocite{kaspi1998}, Kaspi \& Helfand 2002\nocite{kh2002}),  
but it is also difficult to 
know for sure the type of SN which produced a given SNR. 

 

 
Core-collapse SN (in spiral galaxies)  represent about 70-80\% 
of all SN (Cappellaro \& Turatto 2001\nocite{ct2000}). 
Among core-collapse SN the  fractions of different supernova types 
are the following:  
SN IIP (0.3), SN IIL (0.3), SN IIn (0.02), SN 1987A-like (0.15), 
SN Ib/c (0.23) (Chevalier 2003\nocite{c2003}, see details in
Dahlen \& Fransson 1999\nocite{df1999}). 
We note that the distribution of SNae through types  
is different for different types of galaxies,  
it depends on the metallicity and so on the cosmic age.  
As a consequence,  all corresponding distributions (like velocity and  
mass distributions of compact objects) are not universal.  
This fact can lead to different fractions of compact objects and  
different velocity distributions of PSRs in different galaxies.  
 
SN IIP probably originate from 
the most low massive stars which can still produce a SN II.  
SN Ib/c probably are the product of binary evolution (or originate from 
very massive single stars). 
Obviously compact objects which originate from single (isolated) stars 
and compact objects which appeared after disruption of a binary, 
should have different velocities simply due to additional orbital velocity. 
A detailed analysis of this situation was given for example by \cite{it1996}.  
They assumed that no natal kick is added (these authors also assumed that 
PSRs are born only in close binaries, see Tutukov et al 
1984\nocite{tyc1984}). 
As it is clear from their fig.10,  massive binaries can give velocities up to 
1000~km~s$^{-1}$, but the fraction of these high-velocity 
objects in the total number of  compact objects  is small.  
Oppositely AIC can produce low velocity 
 compact stars, but the contribution is also rather 
low: $3 \cdot 10^{-3}$~yr$^{-1}$  (Podsiadlowski et al. 2003\nocite{plp2003}).  
 
 The  Crab pulsar   
is probably associated with SN IIP. In some sense it goes in  line  
with our scenario, as far as compact objects formed from low massive stars 
({\it i.e.} $\sim10-15\, M_{\sun}$) should be low velocity ({\it i.e.} 
$V_\mathrm{t}<250$~km~s$^{-1}$)  HSs and their fraction is $\sim 0.2-0.4$. 
Then the high-velocity component of the velocity distribution should be
related to SN IIL and probably to SN Ib/c.
 
 
\subsection{Predictions and speculations} 
 
There are many predictions connected with the proposed scenario, 
and much more speculations can be done. 
 
As far as fractions of compact objects in the first and in the second 
components of the {\it initial}  
velocity distribution are more or less equal, then within our scenario 
the number of {\it formed} HSs and QSs are also nearly equal.  
On the other hand, high velocity objects quickly leave our Galaxy. 
So, among {\it observable} sources the number of QSs  should be not very large. 
QSs can fill the galactic halo.  
 
If QSs are fast objects, then their fraction in globular clusters should be 
low as far as a typical escape velocity is $<$100~km~s$^{-1}$. 
 
Due to different cooling history of HSs and QSs we can expect a correlation 
in the temperature-velocity relation for young compact objects, {\it i.e.}
temperatures (for the same age) of objects from the first and the second
peaks of the velocity distribution can be different. 
However, the cooling history is mass-dependent both for HSs and QSs, 
so this kind of correlation should not be a clear one. 

 An interesting problem is connected with a possibility of glitches in QSs.
This topic 
is not in the main line of our discussion, so we mention it just briefly.
As far as internal structure of QSs and HSs are different, then
it is reasonable to expect that glitches of the two types of compact objects
have different properties (see an early discussion in Alcock et al.
1986\nocite{afo1986}, and more recent in Horvath 2004\nocite{h2004}).  
At present up to 100 glitches from about 30 PSRs were reported
(see Shemar \& Lyne 1996\nocite{sl1996}, 
Krawczyk et al. 2003\nocite{klg2003} and references therein).
For 8 of these objects in the ATNF catalogue
(Hobbs et al. 2003\nocite{hobbs2003}) transverse
velocities are available. 
Several glitching PSRs with known velocities
can be added from Kaspi (2000\nocite{kaspi1998}).
For the two-component distribution of \cite{acc2002}
contribution of the low velocity component dominate up to $\sim
300$~km~s$^{-1}$, which corresponds to the transverse velocity 
$V_\mathrm{t}\sim 250$~km~s$^{-1}$.
All but two velocity measurments of glitchers are below this limit.
The best known glitchers -- Crab and Vela -- definitely represent
the low-velocity component. 
The most high-velocity PSR among glitchers 
-- 2224+65 --
was reported to have a glitch just once (Shemar \& Lyne 1996\nocite{sl1996})
and detailed properties of this glitch are unknown.
 
In SNRs closer to the centers we expect to see HSs, and QSs 
with higher average velocities should be more displaced (for the same age). 
(See also Zhang et al. 2000\nocite{zxq2000},  
where the authors briefly discuss this topic 
in connection with AXPs and SGRs.) 
 
If there is a delayed quark deconfinement in a single compact object,  
then we can expect (even if the present scenario for the explanation 
of the bimodality for PSR velocity distribution  is not correct) 
to have "young" PSRs without pronounced SNRs. 
Due to stellar contraction and (probably) due to an additional kick 
such objects can be spun-up and warmed, so they can look like 
young {\it neutron stars}.   
In the case of a short time delay between the SN explosion and the 
HS$\rightarrow$QS conversion (so that a SNR formed after the HS 
formation is still visible) there  can be a discrepancy between   
the characteristic PSR age ($\tau=\dot p/2p$), the SNR age,  
and the kinematic age (displacemnet of a compact object from the 
SNR center divided by the transverse velocity).      
 
%
 
The outcome of a SN explosion depends on the metallicity 
(Woosley et al. 2002\nocite{whw2002}). 
The masses of iron cores and oxygen shells,  
the amount of matter fall-back, the rotation rates etc.  
are different for stars with different fraction of heavy elements.  
Metallicity is changing during the lifetime of the Galaxy. 
As a consequence,  one can expect different fractions of newborn QSs 
among compact objects  at different galactic ages.  
 
As far as the formation of a QS from an HS is accompanied by 
the shrinking of the compact object, it can lead  to an increase of the 
stellar magnetic field and to a decrease of the initial spin period. 
However, the details of this process are uncertain. 
\cite{fgw1994} note that in their set of PSRs, the high-velocity ones 
have higher magnetic fields. 
But the results of that paper (mainly the reality of associations 
of PSRs and SNRs) were later doubted. 
Also, the  results mentioned in \cite{kh2002} are against 
any correlation between initial spin and velocity (see data on initial 
spin periods for 7 PSRs in Migliazzo et al. 2002\nocite{m2002}). 
Among these seven PSRs, the transverse velocities are known 
in four cases.  Only one of them has a transverse velocity which 
falls into the second (high-velocity) peak of the velocity distribution. 
This pulsar has the shortest period (even just an upper limit) 
among those reported in (Migliazzo et al. 2002\nocite{m2002}).
We predict that the remaining three with unknown velocities 
should have $V_\mathrm{t}<$~(250--300)~km~s$^{-1}$.  
In general, in our opinion, it is important to check possible links of  
properties connected with the internal structure of compact stars 
with their velocity properties, but not just in a form of correlations: 
it is important to check in which peak of the distribution a given 
compact star falls.  

 
Fall-back process can bring additional correlations between different 
properties  of compact objects (as it was suggested by  
Popov et al. 2002\nocite{popov2002}), 
especially as far as this process can play a key role in shortening 
time of the delay between the SN explosion and 
the  HS$\rightarrow$QS conversion.  
 
There are reasons to suspect, that the magnetic field distribution 
of compact objects is also bimodal because of a significant number of magnetars. 
However, the high-field population is at most 10\%, not comparable with 
the high-velocity fraction of  PSRs.  
 
Nowadays it is more or less clear that not all the {\it neutron stars} 
pass through the stage of active PSR (or at least this stage is very short for them).  
In this sense,  it is interesting to understand if they can be QSs.       
In connection with the proposed scenario,  it is important to note  
that two of the Magnificent seven (seven ROSAT radioquiet NSs) for which  
velocities are known, probably are low-velocity objects:  
$V_\mathrm{t}<(200$~-~$250)$~km~s$^{-1}$.  
However, more precise results will come out soon. 
 

\section{Conclusion} 

The existence of both hadronic and quark stars in comparable 
amount can imply a bimodality of many properties of observed 
compact objects including radio pulsars. 
In this paper,  we discussed  the bimodality of the kick 
velocity distribution of PSRs. 
Within our scenario, the low-velocity peak of the velocity distribution 
corresponds to hadronic stars.  
Quark stars populate the second (high-velocity) part of the distribution.

Such bimodality in the kick velocity distribution can appear 
due to the following reasons: 

\begin{itemize}

\item  QSs can experience two well time separated huge energy releases. 
The first one during a ``normal''  SN explosion, which forms a pure HS. 
The second during a delayed stellar conversion (triggered by quark 
deconfinement) which forms a QS.  
Large second kicks can explain the observed velocity bimodality.  

\item   Even in the case both types of compact stars ({\it i.e.} HSs and QSs)  
are formed promptly during a SN explosion, 
then the kick mechanism can be more efficient for a QS than for a HS.   
For example, we discussed in this paper that the electromagnetic 
rocket kick can be larger for QSs as far as these objects can be born 
with spin periods much shorter than those for HSs. 

\end{itemize}

It is of great interest and importance to compare different 
observational properties of
high- and low-velocity compact stars 
as far as the difference in the initial kick 
can have roots in differences in the internal stellar structure. 
In our opinion,  
it is useful not just to correlate the stellar spatial velocity 
with other physical parameters,  but to look at differences between 
two ensembles of compact stars (two components of the velocity
distribution). Probably these two components
are formed by  different classes of compact objects (QSs and HSs), 
and inside each part correlations can be absent, but between the two parts 
differencies can be significant.

\begin{acknowledgements} 
SP thanks the University of Pisa and INFN for hospitality. 
\end{acknowledgements} 
 


\begin{thebibliography}{} 
 
\bibitem[\protect\citeauthoryear{Alcock et al.}{1986}]{afo1986} 
Alcock, C., Farhi, E., \& Olinto, A. 1986,  ApJ 310, 261 
 
\bibitem[\protect\citeauthoryear{Ardeljan et al.}{2004}]{abm2003} 
Ardeljan, N.V., Bisnovatyi-Kogan, G.S.,  Kosmachevskii, K. V., 
\& Moiseenko, S.G. 2004,  
Astrophysics (English translation of Astrofizika) 47,  37 
 
\bibitem[\protect\citeauthoryear{Arzoumanian et al.}{2002}]{acc2002} 
Arzoumanian, Z., Chernoff, D.F., \& Cordes, J.M. 
2002, ApJ 568, 289 
 

\bibitem[\protect\citeauthoryear{Baldo et al.}{1997}]{bbb97} 
Baldo, M.,  Bombaci, I., \& Burgio, G.F. 1997, A\&A, 328, 274

\bibitem[\protect\citeauthoryear{Benvenuto\& Horvath}{1989}]{bh1989} 
Benvenuto, O.G., \& Horvath, J.E. 1989, Phys. Rev. Lett. 63, 716 
 
\bibitem [\protect\citeauthoryear{Berezhiani et al.}{2003}]{bbd2003} 
Berezhiani, Z.,  Bombaci, I., Drago, A.,  Frontera, F., \& Lavagno, A. 2003,  
ApJ 586, 1250 
 

\bibitem[\protect\citeauthoryear{Bodmer}{1971}]{bod71} 
Bodmer, A.R.  1971, Phys. Rev. D, 4, 1601

\bibitem [\protect\citeauthoryear{Bombaci \& Datta}{2001}]{bd2000} 
Bombaci, I., \& Datta, B. 2000, ApJ 530, L69 

\bibitem [\protect\citeauthoryear{Bombaci et al.}{2004}]{bpv2004} 
Bombaci, I., Parenti, I., \& Vida\~na, I. 2004,   astro-ph/0402404 

\bibitem [\protect\citeauthoryear{Brisken et al.}{2003}]{bfg2003} 
Brisken, W.F., Fruchter, A.S., Goss, W.M., Herrnstein, R.S., \& Thorsett, S.E.  
2003,  AJ 126, 3090 
 
\bibitem[\protect\citeauthoryear{Burrows et al.}{2003}]{bom2003} 
Burrows, A., Ott, C.D., \& Meakin, C. 2003,  astro-ph/0309684 
 
\bibitem[\protect\citeauthoryear{Cappellaro \& Turatto}{2001}]{ct2000} 
Cappellaro, E., \& Turatto, M. 2001, 
in "The influence of binaries on stellar population studies", 
Dordrecht: Kluwer Academic Publishers, Astrophysics and space science
library vol. 264, p.199 
[astro-ph/0012455] 
 
 
\bibitem[\protect\citeauthoryear{Chevalier}{2003}]{c2003} 
Chevalier, R.A. 2003, in Proc. IAU Symp. 218, p.97
[astro-ph/0310730] 
 
\bibitem[\protect\citeauthoryear{Colpi \& Wasserman}{2002}]{cw2002} 
Colpi, M., \& Wasserman, I. 2002, ApJ 581, 1271  
 
 
\bibitem[\protect\citeauthoryear{Dahlen \& Fransson}{1999}]{df1999} 
Dahlen, T., \& Fransson, C. 1999, A\&A 350, 349 

\bibitem[\protect\citeauthoryear{Dey et al.}{1998}]{dey98} 
Dey, M., Bombaci, I.,  Dey, J., Ray, S., \& Samanta, B.C. 1998, 
                                   Phys. Lett. B, 438, 123; erratum 1999, Phys. Lett. B, 467, 303

\bibitem[\protect\citeauthoryear{Frail et al.}{1994}]{fgw1994} 
Frail, D.A., Goss, W.M., \& Whiteoak, J.B.Z. 1994, ApJ 437, 781 


\bibitem[\protect\citeauthoryear{Glendenning}{1992}]{gle92} 
Glendenning, N.K 1992, Phys. Rev. D, 46, 1274

\bibitem[\protect\citeauthoryear{Glendenning}{1996}]{gle96}   
Glendenning,  N.K., 1996,  Compact Stars: Nuclear Physics, 
                               Particle Physics, and General Relativity, Springer Verlag 

\bibitem[\protect\citeauthoryear{Green \& Stephenson}{2003}]{gs2003} 
Green, D.A., \& Stephenson, F.R. 2003, in
"Supernovae and Gamma-Ray Bursters", ed. K. Weiler. Lecture Notes in
Physics, vol. 598. Berlin, New York: Springer,  p.7  
[astro-ph/0301603]  

\bibitem[\protect\citeauthoryear{Haensel et al.}{1986}]{hae86} 
Haensel, P., Zdunik, J.L., \& Schaefer, R. 1986, A\&A, 160, 121

\bibitem[\protect\citeauthoryear{Harrison \& Tademaru}{1975}]{ht1975} 
Harrison, E.R., \& Tademaru, E. 1975, ApJ 201, 447 
 
 
\bibitem[\protect\citeauthoryear{Hobbs et al.}{2003}]{hobbs2003} 
Hobbs G. et  al. 2003, astro-ph/0309219 
 
\bibitem [\protect\citeauthoryear{Hong et al.}{2001}]{hhs2001} 
Hong, D.K., Hsu, S.D.H., \& Sannino, F. 2001, 
 Phys.Lett. B516, 362 

\bibitem [\protect\citeauthoryear{Horvath}{2004}]{h2004}
Horvath, J.E. 2004, astro-ph/0404324
 
\bibitem[\protect\citeauthoryear{Huang et al.}{2003}]{hdl2003} 
Huang, Y.F., Dai, Z.G., Lu, T., Cheng, K.S., \& Wu, X.F. 2003, 
ApJ 594, 919 
 
\bibitem[\protect\citeauthoryear{Iben \& Tutukov}{1996}]{it1996} 
Iben, I., \& Tutukov, A.V. 1996, ApJ 456, 738  
 
\bibitem[\protect\citeauthoryear{Imshennik}{1992}]{i1992} 
Imshennik, V.S. 1992, PAZh 18, 489 
 
\bibitem[\protect\citeauthoryear{Imshennik \& Nadezhin}{1992}]{in1992} 
Imshennik, V.S., \& Nadezhin, D.K., 1992, PAZh 18, 79 
 
 
\bibitem[\protect\citeauthoryear{Imshennik \& Ryazhskaya}{2004}]{ir2004} 
Imshennik, V.S., \& Ryazhskaya, O.G., 2004, Astron. Lett. 29, 
831 [astro-ph/0402191] 
 
 
 
\bibitem[\protect\citeauthoryear{Kaspi}{2000}]{kaspi1998} 
Kaspi, V. 2000, in "Pulsar Astronomy~--~2000 and Beyond", ASP Conference
Series, vol. 202; Proc. IAU Coll. 177,
ed. M. Kramer, N. Wex, \& N. Wielebinski, p. 485 
[astro-ph/9803026] 
 
\bibitem[\protect\citeauthoryear{Kaspi \& Helfand}{2002}]{kh2002} 
Kaspi, V.M., \& Helfand, D.J. 2002, in "Neutron Stars in Supernova Remnants",
ASP Conference Series, vol. 271,
ed. P.O. Slane, \& B.M. Gaensler, San Francisco: ASP,
p.3 
[astro-ph/0201183] 
 
 
\bibitem[\protect\citeauthoryear{Krawczyk et al.}{2003}]{klg2003} 
Krawczyk, A., Lyne, A.G., Gil, J.A., \& Joshi, B.C. 2003, 
MNRAS 340 1087 
 
\bibitem[\protect\citeauthoryear{Lai}{2003}]{lai2002} 
Lai, D. 2003, in "Radio Pulsars", ASP Conference Proc., vol. 302,
 ed. M. Bailes, D.J. Nice, \& S.E. Thorsett, San
Francisco: Astron. Soc. of the Pacific, 
p.307 
[astro-ph/0212140] 
 
\bibitem[\protect\citeauthoryear{Lai et al.}{2001}]{lcc2001} 
Lai, D., Chernoff, D.F., \& Cordes, J.M. 2001, ApJ 549, 1111 
 


\bibitem[\protect\citeauthoryear{Li et al.}{1999}]{li99a}    
 Li, X.--D., Bombaci, I., Dey, M., Dey J., \& van den Heuvel, E.P.J. 
                                   1999a, Phys. Rev. Lett., 83, 3776  

\bibitem[\protect\citeauthoryear{Lorimer et al.}{1997}]{lbh1997} 
Lorimer, D.R., Bailes, M., \& Harrison, P.A. 1997, MNRAS 289, 592 
 
\bibitem[\protect\citeauthoryear{Lyne \& Lorimer}{1994}]{ll1994} 
Lyne, A.G., \& Lorimer, D.R. 1994, Nature  369, 127 
 
\bibitem[\protect\citeauthoryear{Lyne et al.}{1998}]{lyne1998} 
Lyne, A.G. et al. 1998,  MNRAS 295, 743 
 
 
\bibitem[\protect\citeauthoryear{Madsen}{1998}]{m1998} 
Madsen, J. 1998, Phys. Rev. Lett. 81, 3311 
 
\bibitem[\protect\citeauthoryear{Migliazzo et al.}{2002}]{m2002} 
Migliazzo, J.M., Gaensler, B.M. et al. 2002, ApJ 567, L141 
 
\bibitem[\protect\citeauthoryear{Moiseenko et al.}{2003}]{mba2003} 
Moiseenko, S.G., Bisnovatyi-Kogan, G.S., \& Ardeljan, N.V. 2003, in
Proc. of IAU Coll. 192, "Supernovae: 10 years after 1993j" 
astro-ph/0310142  
 

\bibitem[\protect\citeauthoryear{Nucastro et al.}{2001}]{n2001}
Nicastro, L., Nigro, F., D'Amico, N., Lumiella, V., \& Johnston, S. 2001,
A\&A 368, 1055 

 
 
\bibitem[\protect\citeauthoryear{Pethich \& Ravenhall}{1995}]{pr95} 
Pethick, C.J., \& Ravenhall, D.G. 1995, Ann. Rev. Nucl. Part. Sci. 45, 429   
 
\bibitem[\protect\citeauthoryear{Podsiadlowski et al.}{2003}]{plp2003} 
Podsiadlowski, Ph., Langer, N., Poelarends, A.J.T., Rappaport, 
S., Heger, A., \& Pfahl, E. 2003,  
astro-ph/0309588 
 
\bibitem[\protect\citeauthoryear{Popov et al.}{2002}]{popov2002} 
Popov, S.B., et al. 2002, in Proc. of the Third International
Sakharov Conference on Physics, ed. A.Semikhatov et al. , Scientific
World: Moscow, p 420 
[astro-ph/0210688] 
 
\bibitem[\protect\citeauthoryear{Postnov \& Prokhorov}{1998}]{pp1998} 
Postnov, K.A., \& Prokhorov, M.E. 1998, Astronomy Letters 24, 568 


\bibitem[\protect\citeauthoryear{Prakash et al.}{1997}]{prak97}  
 Prakash, M., Bombaci, I., Prakash, M., Ellis, P.J.,  Lattimer, J.M., 
                                    \&   Knorren, R. 1997,  Phys. Rep. 280, 1  

\bibitem[\protect\citeauthoryear{Romani}{2004}]{r2004}
Romani, R.W. 2004, astro-ph/0404100 

\bibitem[\protect\citeauthoryear{Salmonson \& Wilson}{1999}]{sw1999} 
Salmonson, J.D., \& Wilson, J.R. 1999, ApJ 517, 859 
 
\bibitem[\protect\citeauthoryear{Shemar \& Lyne}{1996}]{sl1996} 
Shemar, S.L., \& Lyne, A.G. 1996, MNRAS 282, 677 
 
\bibitem[\protect\citeauthoryear{Spruit \& Phinney}{1998}]{sp1998} 
Spruit, H., \& Phinney, E.S. 1998,  Nature 393, 139 
 
\bibitem[\protect\citeauthoryear{Terazawa}{1979}]{ter79} 
Terazawa, H. 1979, INS-Report, 336 (INS, Univ. of Tokyo); 
                                          1989, J. Phys. Soc. Japan, 58, 3555; 
                                         1989, J. Phys. Soc. Japan, 58, 4388


 
\bibitem[\protect\citeauthoryear{Tutukov et al.}{1984}]{tyc1984} 
Tutukov, A.V., Chugai, N.N., \& Yungelson, L.R. 1984, 
Sov. Astron. Lett. 10, 244 
 
\bibitem[\protect\citeauthoryear{Van Dyk et al.}{2003}]{vlf2002} 
Van Dyk, S.D., Li, W., \& Filippenko, A.V. 2003, PASP 115, 1 
 

\bibitem[\protect\citeauthoryear{Witten}{1984}]{witt84} 
Witten, E. 1984, Phys. Rev. D, 30, 272

\bibitem[\protect\citeauthoryear{Woosley et al.}{2002}]{whw2002} 
Woosley, S.E., Heger, A., \& Weaver, T.A. 2002,   
Reviews of modern physics 74, 1015 
 
 
\bibitem[\protect\citeauthoryear{Zhang et al.}{2000}]{zxq2000} 
Zhang, B., Xu, R.X., \& Qiao, G.J. 2000, ApJ 545, L127 
 
 
\end{thebibliography}
\end{document}